\documentclass[pdflatex, sn-nature]{sn-jnl}
\usepackage{graphicx}
\usepackage{multirow}
\usepackage{amsmath, amssymb, amsfonts}
\usepackage{amsthm}
\usepackage{mathrsfs}
\usepackage[title]{appendix}
\usepackage{xcolor}
\usepackage{textcomp}
\usepackage{manyfoot}
\usepackage{booktabs}
\usepackage{setspace}
\usepackage{hyperref}
\hypersetup{colorlinks=true, linkcolor=blue, citecolor=blue, urlcolor=blue}

\theoremstyle{thmstyleone}

\theoremstyle{thmstyletwo}

\theoremstyle{thmstylethree}

\begin{document}

\title[Stimulated Emission from Boson Clouds]{Stimulated Emission from Boson Clouds}

\author[1]{\fnm{Yu} \sur{An}\email{anulinkarc@gmail.com}}

\author[1]{\fnm{Xian-Hui} \sur{ Ge}}\email{gexh@shu.edu.cn}

\author[2]{\fnm{Yun-Gui} \sur{Gong}\email{gongyungui@nbu.edu.cn}}

\author[3]{\fnm{Yun-Long} \sur{Zhang}}\email{zhangyunlong@nao.ac.cn}

\affil[1]{\orgdiv{Department of Physics, College of Sciences}, \orgname{Shanghai University}, \orgaddress{\street{99 Shangda Road}, \city{Shanghai}, \postcode{200444}, \state{Shanghai}, \country{China}}}
\affil[2]{\orgdiv{Department of Physics, Institute of Fundamental Physics and Quantum Technology}, \orgname{Ningbo University}, \orgaddress{\street{818 Fenghua Road}, \city{Ningbo}, \postcode{315211}, \state{Zhejiang}, \country{China}}}
\affil[3]{\orgdiv{National Astronomical Observatories}, \orgname{Chinese Academy of Sciences}, \orgaddress{\street{Datun Road A20}, \city{Beijing}, \postcode{100101}, \state{Beijing}, \country{China}}}

\abstract{\quad Gravitational-waves from astrophysical sources are characterized by their extreme faintness, which remains a primary obstacle for both current and next generation detectors. While rotating black holes dressed in superradiant clouds of ultralight bosons are recognized as promising probes of physics beyond the Standard Model, their capacity to actively emit and modulate gravitational radiation remains largely unexamined. Here we demonstrate that these gravitational atoms can function as natural amplifiers of gravitational-waves via a stimulated emission mechanism analogous to astrophysical masers. By formalizing the interaction between the bosonic cloud and an ambient stochastic gravitational-wave background, we establish the rigorous selection rules and threshold conditions that govern this amplification. Our analysis reveals that the emission rate depends critically on the boson mass, potentially yielding an enhancement of several orders of magnitude over spontaneous processes. For representative mass ranges, these amplified signals bridge the sensitivity gap between ground-based interferometers and pulsar timing arrays. These findings suggest that superradiant clouds can effectively boost previously undetectable signals, offering a novel observational frontier for exploring ultralight fields and the Kerr spacetime environment. 
}


\maketitle
\section{Introduction}\label{sec1}

\quad As ripples in spacetime predicted by general relativity, gravitational-waves (GWs) are characterized by their extreme faintness, posing a formidable challenge for current detection technologies, especially when originating from distant or low-luminosity astrophysical sources \citep{Abbott2016}. Overcoming this sensitivity threshold is a prerequisite for the continued advancement of gravitational-wave astronomy. We propose that gravitational atoms systems comprising Kerr black holes enveloped by superradiantly grown clouds of ultralight scalar bosons can function as natural GW amplifiers through a stimulated emission mechanism directly analogous to the operation of atomic lasers. Unlike spontaneous emission, stimulated emission in a bound-state system occurs when an external wave, with a frequency matching the energy transition between two discrete levels, induces a particle to drop to a lower-energy state. In this gravitational framework, the rotating black hole acts as the atomic nucleus, while the dense boson cloud occupies hydrogenic bound states, fulfilling the fundamental requirements for coherent GW amplification. 

The theoretical construction of a gravitational atom draws a parallel to the hydrogen atom, where the black hole's gravitational potential binds the ultralight boson field. This synergy stems from the superradiance phenomenon first identified by Misner in 1972 which allows scalar waves incident on a black hole’s ergoregion to extract rotational energy and undergo amplification. This concept was subsequently extended to the "black hole bomb" mechanism, wherein repeated superradiant scattering leads to the accumulation of substantial energy within a localized boson cloud \citep{Press1972}. In 2010, Arvanitaki et al. suggested that ultralight bosons, such as axions, could form stable clouds due to their negligible coupling to Standard Model particles. Given the superradiance condition $\omega \leq m \Omega_H$, the boson mass is inversely coupled to the black hole mass, facilitating the formation of bound states characterized by hydrogen-like wavefunctions \citep{Arvanitaki2011, Arvanitaki2010, Brito2020}. While the theoretical foundations were established by Detweiler \citep{Detweiler1980}, more recent studies by Brito et al. have highlighted the potential for these systems to generate detectable GW signals for observatories such as LIGO and LISA \citep{Brito2015, Brito20172}. Notably, while spin-1/2 particles do not exhibit classical superradiance, as demonstrated by Unruh  and Chandrasekhar, scalar and vector fields are readily amplified in the Kerr spacetime\citep{Unruh1973, Toth2016, Chandrasekhar1976, Iyer1978, Dolan2015}. 

Recent empirical constraints, including black hole spin measurements analyzed by Cardoso et al. , suggest that ultralight bosons interact only minimally, if at all, with Standard Model particles\citep{Cardoso2018}. This isolation implies that gravitational interactions are the dominant, if not exclusive, channel for transition phenomena within gravitational atoms. Consequently, the detection and characterization of these bosons which remain prime candidates for cold dark matter \cite{Baryakhtar2015} rely heavily on our ability to interpret their gravitational signatures. Probing these ultralight fields is thus a cornerstone of modern fundamental physics. 

Given that gravitational interactions govern these systems, GWs serve as the primary external driver for inducing energy level transitions. Inspired by the principles of quantum optics, we hypothesize that an incident GW with frequency $k_0$, satisfying the resonance condition $E_{nlm} = k_0 + E_{n'l'm'}$, can trigger a stimulated transition of bosons from excited states to lower energy levels, resulting in the emission of a coherent GW. Considering that these clouds can sequester up to $10^{-1}$ of the black hole's mass, the stimulated process may amplify faint GW signals by multiple orders of magnitude. While previous models, such as those by Zhang and Yang and Baumann et al. , explored gravitational excitations in binary systems, they focused primarily on absorption and tidal resonances\citep{Zhang2020, Baumann2022, Baumann2019}. In this work, we shift the focus to the emission sector, demonstrating that stimulated emission in gravitational atoms represents a viable and potentially transformative mechanism for observing ultralight particles and the intricate environments of spinning black holes. 

The stochastic gravitational-wave background (SGWB), predominantly generated by unresolved populations of compact binaries, supplies a continuous frequency spectrum\citep{Maggiore2000, Allen1999, Romano2017}. This broadband radiation naturally provides the requisite resonant frequencies to trigger stimulated emission within the superradiant cloud. Such a process would imprint distinct spectral peaks atop the stochastic continuum, offering a novel observational signature. To formalize this amplification mechanism, we adopt natural units ($G_{N} = \hbar = c = 1$) with the metric signature $(-, +, +, +)$. The background spacetime of the rotating black hole is described by the Kerr metric, which in Boyer-Lindquist coordinates reads:
\begin{equation}
ds^{2} = -\frac{\Delta}{\rho^{2}}(dt - a \sin^{2}\theta d\varphi)^{2} + \frac{\rho^{2}}{\Delta}dr^{2} + \rho^{2}d\theta^{2} + \frac{\sin^{2}\theta}{\rho^{2}}\left[adt - (r^{2} + a^{2})d\varphi\right]^{2}, 
\end{equation}
where $M$ and $J$ represent the mass and angular momentum of the black hole, respectively, and the spin parameter is defined as $a = J/M$. The metric functions are $\Delta = r^{2} - 2Mr + a^{2}$ and $\rho^{2} = r^{2} + a^{2}\cos^{2}\theta$. The inner and outer event horizons are located at $r_{\pm} = M \pm \sqrt{M^{2} - a^{2}}$, with the outer horizon rotating at an angular velocity of $\Omega_{+} = \frac{a}{2Mr_{+}}$. 

\section{Mechanism of Gravitational-Wave Stimulated Emission}\label{sec2}\subsection{Structure of the Gravitational Atom}\quad The gravitational atom is modelled as a central Kerr black hole dressed by a macroscopic cloud of ultralight scalar bosons. This cloud is dynamically generated via superradiance, a mechanism through which the bosonic field extracts rotational energy and angular momentum from the black hole\citep{Detweiler1980}. Assuming spin-0 particles, the dynamics of the boson field $\phi(t, \vec{r})$ are governed by the massive Klein-Gordon equation in the Kerr spacetime:
\begin{equation}g_{\mu\nu}\nabla^{\mu}\nabla^{\nu}\phi(t, \vec{r}) - m_b^2 \phi(t, \vec{r}) = 0, \end{equation}
where $g_{\mu\nu}$ is the background Kerr metric and $m_b$ is the bare mass of the ultralight boson. Efficient superradiant growth requires the gravitational fine-structure constant, $\alpha = M m_b$ (in natural units), to be of order unity. This resonance condition dictates that the most violently amplified boson mass scales inversely with the mass of the host black hole. Analogous to the electronic orbitals in a hydrogen atom, the boson cloud occupies discrete, quasi-bound states. The spatial profile of these states typically extends across several gravitational radii ($r_g = M$) of the central black hole\citep{Baumann2022}. Both numerical relativistic simulations and pseudo-Newtonian treatments confirm that this superradiant condensate can sequester up to $\sim 10\%$ of the initial black hole mass, maintaining dynamical stability over multiple superradiant e-folding times\citep{Detweiler1980}. Crucially, because the fundamental and most unstable modes are localized at distances $r \sim \mathcal{O}(10r_g) $, the bulk of the boson cloud resides in the weakly curved region of spacetime\cite{Arvanitaki2011}. This spatial segregation rigorously justifies the application of a local weak-field approximation when evaluating the stimulated emission processes in the subsequent analysis.

Stimulated emission of gravitational-waves is fundamentally a microscopic process governed by the interaction between the ambient GW field and the quantized ultralight boson cloud. Given the macroscopic spatial extent of the condensate, the dominant interaction occurs in regions where the background curvature is sub-dominant. This local weak-field approximation is rigorously justified in the long-wavelength regime, characterized by the gravitational fine-structure constant:\begin{equation}\alpha = \frac{r_g}{\lambda_c} = \frac{G M m_b}{\hbar c} = M m_b \ll 1, \end{equation}where $\lambda_c = \hbar / (m_b c)$ is the Compton wavelength of the boson. Near the parameter space where superradiance is most efficient ($\alpha \approx 0. 1$)\citep{Brito2015}, the background metric can be perturbatively expanded around a flat spacetime. This non-relativistic reduction yields a Schrödinger-like equation that dictates the spectrum of the quasi-bound states (See Supplementary Material 4. 1). In the long-wavelength limit where the dimensionless parameter $\alpha = M m_b \ll 1$, the dynamics of the ultralight boson cloud simplify to a non-relativistic quantum system governed by hydrogen-like wavefunctions $\psi_{nlm}(t, \vec{r}) = R_{nl}(r) Y_{lm}(\theta, \varphi) e^{-i (\omega_{nlm} - m_b) t}$. The quasi-bound states of this gravitational atom are classified by the principal, azimuthal, and magnetic quantum numbers ($n, l, m$) with eigenenergies well-approximated by the relation $E_{nlm} \approx m_b \left(1 - \frac{\alpha^2}{2n^2}\right)$ \citep{Baumann2022}, where the gravitational coupling $\alpha^2$ naturally replaces the electromagnetic fine-structure constant. Given that the gravitational fine-structure constant \(\alpha\) remains small during the most vigorous superradiance process, and since the instability rate \(\Gamma_{nlm}\) (see Eq. (14)) in the wavefunction representation manifests as a convergent growth term of the boson cloud and scales as \(\alpha^{4l+5}\), it consequently constitutes a higher-order small quantity relative to the bound-state energies. Therefore, the contribution of the instability rate, after the black hole superradiance has proceeded for a sufficiently long time, manifests itself through the boson number and angular momentum of the cloud outside the black hole \citep{Baumann2022, Baumann2019}. This enduring stability enables a mechanism strictly analogous to the operation of an atomic laser. An incident gravitational-wave can resonantly interact with the superradiant cloud and trigger a downward transition between the bosonic orbitals, stimulating the emission of coherent gravitational radiation. By operating in the maximal superradiance regime ($\alpha = 0. 1$) for our subsequent analysis, we can safely truncate higher-order relativistic corrections and rigorously compute the transition amplitudes driven by this external perturbation. 

The rate of stimulated emission is fundamentally dictated by the transition matrix elements between the quasi-bound states, mediated by an interaction Hamiltonian that couples the scalar condensate to the perturbing tensor field. Crucially, an incident gravitational-wave (GW) satisfying the resonant frequency condition induces a downward transition, yielding coherent gravitational radiation that is strictly phase- and polarization-locked to the seed wave. Because the superradiant cloud harbors a macroscopic number density of bosons extending over several gravitational radii\citep{Arvanitaki2011}. This collective coherence allows the gravitational atom to function as a highly efficient, natural GW amplifier, primed to boost both transient signals from proximate compact binaries and the ambient stochastic background. The observational consequences of this resonant amplification are physically distinct. Because stimulated emission strictly requires the incident wave to match the exact energy difference between specific atomic transitions, the amplification does not simply produce a uniform broadband enhancement of the overall background signal. Instead, the process extracts energy from the boson cloud to generate highly localized monochromatic peaks superposed on the continuous stochastic spectrum. Depending on the rest mass of the ultralight boson, these distinct spectral lines will fall directly within the targeted frequency bands of current and future interferometric detectors such as LIGO, Virgo, KAGRA, the Einstein Telescope, and LISA \cite{Brito2017, Baryakhtar2015}. The most prominent of these signatures would be the narrowband emission lines, which emerge directly from the discrete energy differences of specific $(n, l, m) \to (n', l', m')$ transitions. These monochromatic, coherent injections would not only offer a pristine observational probe into the mass spectrum of ultralight bosons, but also establish an unprecedented astrophysical laboratory for precision tests of general relativity in the near-horizon regime of Kerr black holes. 

\subsection {Amplification of Gravitational Atom Transitions Driven by a Stochastic Gravitational-Wave Background}
To formalize the transition dynamics, we evaluate the interaction between the gravitational atom and an ambient stochastic gravitational-wave background a persistent feature of dense astrophysical environments, such as galactic centers or regions populated by compact binaries. Working within a field-theoretic framework, this interaction is captured by a Lagrangian density that couples the ultralight scalar field to the external tensor metric perturbations, $h_{\mu\nu}$, evaluated in the transverse-traceless gauge. As justified by the local weak-field approximation in the long-wavelength limit, the background Kerr geometry across the macroscopic volume of the cloud is effectively treated as a flat Minkowski spacetime. This robustly simplifies the covariant derivatives to standard partial derivatives. Consequently, the leading-order interaction Lagrangian density takes the form $\mathcal{L}_I = \frac{1}{2} h_{\mu\nu} \partial^{\mu} \phi \partial^{\nu} \phi$. Passing to the Hamiltonian picture, this yields the interaction Hamiltonian density:\begin{equation}\mathcal{H}_{I} = -\frac{1}{2} h_{\mu\nu} \partial^{\mu} \phi \partial^{\nu} \phi. \end{equation}It is this interaction Hamiltonian that acts as the primary perturbative operator, governing the stimulated transitions between the discrete quasi-bound states of the superradiant condensate. 

Before formalizing the transition dynamics, it is crucial to establish the physical distinction between spontaneous and stimulated emission within the context of a gravitational atom. Spontaneous emission occurs when a boson decays to a lower energy state by emitting a graviton into the vacuum. Because this process relies entirely on the inherent gravitational coupling, the vacuum decay is exponentially suppressed. In contrast, stimulated emission is a resonantly driven process. When an ambient gravitational field, most notably the stochastic gravitational-wave background, permeates the condensate, frequency components that exactly match the atomic energy gap can forcefully perturb the system. This resonant interaction triggers the atom to emit gravitational radiation that shares the precise energy, phase, and polarization of the incident wave, significantly accelerating the overall decay process. We compute the probability of this stimulated transition using the S-matrix formalism under the adiabatic approximation, assuming the interaction persists from the infinite past to the infinite future. Expanding the S-matrix via a Dyson series yields the first-order transition amplitude from an initial excited level $|nlm\rangle$ to a final state $|n'l'm'; k_z, -2\rangle$. Taking the specific transition from $|322\rangle$ to $|100; k_z, -2\rangle$ as a primary example, applying Wick's theorem ensures that only normal-order terms contribute due to strict selection rules. The critical feature enabling this dynamic is the resonance condition $E_{nlm} = k_0 + E_{n'l'm'}$. For the aforementioned states, the energy of the incident gravitational-wave $k_0$ must precisely match the internal energy difference, giving $k_0 = E_{322} - E_{100} = \frac{4}{9} m_b \alpha^2$. Evaluating these dynamics through Fermi's golden rule reveals a huge hierarchy between spontaneous and stimulated rates. For a benchmark ultralight boson mass of $m_b = 4\times 10^{-46}\, \text{kg}$ forming a macroscopic condensate of $N = 10^{75}$ bosons, the collective spontaneous decay rate remains entirely negligible ($\Gamma_{\text{sp}}^{(\text{tot})} \sim 1. 15 \times 10^{-1}\, \text{s}^{-1}$). However, the presence of an ambient SGWB with a characteristic strain amplitude of $h_c(f) = 10^{-24}$ drives the collective stimulated emission to a theoretical upper limit of $\Gamma_{\text{st}} \sim 1.6 \times 10^{28}\, \text{s}^{-1}$. This astounding $\frac{\Gamma_{st}}{\Gamma_{sp}}\sim 10^{29}$ amplification factor steepens even further for lighter boson candidates.    While these transition rates appear exceptionally high, they do not trivially translate to macroscopic gravitational-wave luminosities. The total energy released per coherent cascade is fundamentally moderated by the infinitesimally small mass quantum $E \sim m_b$. Furthermore, these upper limits represent an idealized scenario wherein the entire macroscopic population $N$ simultaneously occupies a specific resonant state. In realistic astrophysical environments, the superradiant cloud will populate a spectrum of quasi-bound levels, naturally reducing the effective fractional occupation. Nevertheless, the staggering discrepancy between the spontaneous and stimulated decay times unambiguously demonstrates that external backgrounds can act as a potent trigger, amplifying otherwise dormant signatures into the observable regime. 

\begin{figure}[!t] 
	\centering
	\includegraphics[scale=0.2]{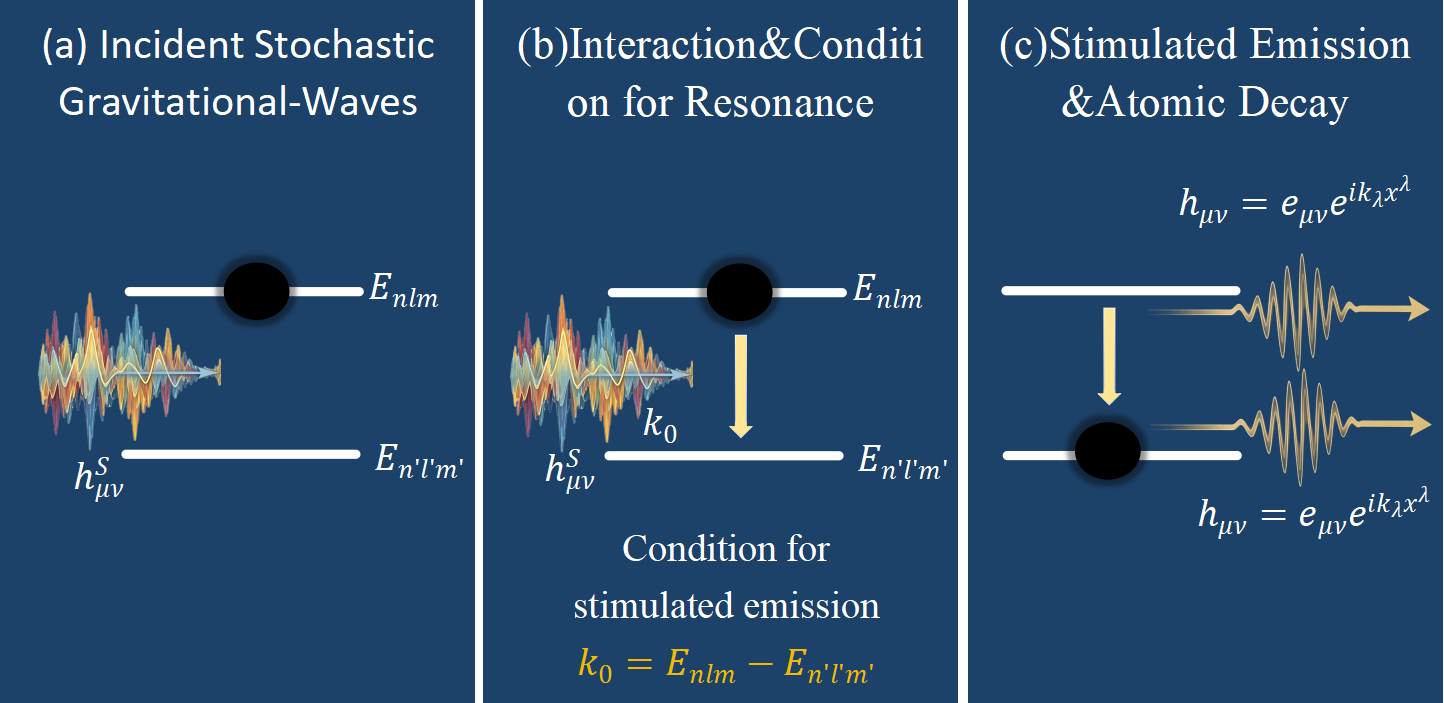}
	\caption{The image illustrates the process of stimulated emission of random gravitational-waves. Specifically, (a) shows an ultralight boson in a gravitational atom, initially in the \(E_{nlm}\) state, being influenced by random gravitational-waves. (b) depicts the ultralight boson transitioning to a lower energy level \(E_{n'l'm'}\) under the induction of a random partial wave with a specific frequency and energy, where the energies satisfy \(E_{nlm} = k_0 + E_{n'l'm'}\). (c) shows that, under the induction of a specific gravitational-wave, the ultralight boson transitions to a lower energy level while simultaneously emitting an additional gravitational-wave that is completely identical to the incident wave. }
	\label{fig2}
\end{figure}

To contextualize these transition rates observationally, we translate the emission luminosities into characteristic strain amplitudes. Assuming a luminosity distance of $d_L = 10^9\, \text{ly}$ ($\sim 300\, \text{Mpc}$) from the gravitational atom, our calculations indicate that the stimulated emission yields a strain of $h \sim 10^{-47}$, compared to a baseline of $h \sim 10^{-63}$ for the spontaneous counterpart. While the stimulated mechanism robustly delivers a 16-order-of-magnitude enhancement over spontaneous decay, the absolute signal strength driven by a diffuse stochastic trigger remains deeply sub-threshold for contemporary and planned interferometric detector networks. Importantly, our obtained results only reflect the conclusions drawn when taking stochastic gravitational-waves as the source. However, the amplification effect for gravitational-waves is genuine. To obtain conclusions that better align with current detection ranges, one may replace the source with a gravitational-wave source capable of producing a larger characteristic strain. If the incident wave is instead sourced by a proximate, high strain dynamical event such as a companion black hole inspiraling through the boson cloud in a binary gravitational atom system the amplitude of the driving field would escalate dramatically. In such configurations, the intense localized gravitational radiation from the companion would act as a powerful resonant trigger, potentially driving the stimulated emission into a highly luminous, observable regime. Therefore, rather than a definitive limit, these results establish coherent amplification as a robust physical mechanism, spotlighting binary gravitational atoms as a highly compelling frontier for future phenomenological studies. 

The explicit focus on the $|3, 2, 2\rangle$ excited state is strictly dictated by the selection rules governing gravitational radiation. Because the emitted graviton carries a spin of two, leading-order transitions fundamentally require a quadrupolar shift in the quantum numbers, satisfying $\Delta l \ge 2$ and $\Delta m = \pm 2$. Consequently, the transition from the superradiantly populated $|3, 2, 2\rangle$ state to the $|1, 0, 0\rangle$ ground state serves as the dominant, lowest-energy channel strictly permitted for coherent single-graviton emission. While a realistic superradiant cloud spans a thermal distribution of various quasi-bound levels, the continuous and highly stable nature of the astrophysical SGWB ensures a persistent driving force. Under these conditions, the incident metric perturbations act collectively across the entire spatial volume of the condensate. This means the field does not merely scatter off individual bosons but induces a phase-coherent, macroscopic perturbation, ensuring that even a fractional occupation of the proper resonant state is sufficient to drive the massive amplification rates described above. 

Building on these strict quadrupolar selection rules, the gravitational atom effectively acts as a highly selective quantum filter when immersed in a broadband stochastic background. Only the specific frequency components of the continuous spectrum that simultaneously satisfy the exact energy resonance condition and the required angular momentum transfer are preferentially amplified. This resonant filtration fundamentally sculpts the output spectrum, guaranteeing the emergence of sharp, discrete emission lines embedded atop an otherwise continuous stochastic foreground. 

Furthermore, the phase coherence of these discrete lines is exceptionally well-protected. In standard astrophysical masers, phase coherence is ultimately degraded by collisional dephasing and thermal fluctuations within the baryonic medium. By contrast, the negligible coupling of ultralight bosons to Standard Model particles ensures that the superradiant condensate remains virtually immune to environmental scattering, even in the presence of dense and turbulent accretion disks. Coupled with the intrinsically weak interaction cross-section of gravitational-waves, this absolute isolation profoundly suppresses any collisional dephasing. Therefore, the stimulated emission is not only narrowly peaked in frequency but is characterized by an exceptionally long coherence time, allowing these systems to function as extraordinarily stable, long-lived gravitational-wave beacons. 

The macroscopic realization of coherent stimulated emission within these extreme astrophysical environments is theoretically robust. A compelling phenomenological analogue is found in astrophysical masers \citep{Elitzur1992, Gray2012}, which prove that macroscopic radiation amplification can spontaneously emerge in nature without explicit optical cavities. As summarized in Table 1, however, the underlying mechanics of the gravitational atom diverge fundamentally from its electromagnetic counterpart. Dictated by intrinsic resonance conditions, the gravitational mechanism maps specific ultralight boson masses to distinct observational frequency bands. For instance, a mass of $m_b = 4 \times 10^{-46}\, \mathrm{kg}$ corresponds to a gravitational-wave frequency of $f \approx 241. 47\, \mathrm{Hz}$, while $m_b = 10^{-58}\, \mathrm{kg}$ scales down to $f \approx 240\, \mathrm{pHz}$. Furthermore, a critical departure lies in the thermodynamic requirements for amplification. While conventional masers strictly demand complex environmental pumping to maintain a population inversion and overcome thermal absorption, the superradiant condensate circumvents this bottleneck entirely. Governed by Bose-Einstein statistics, the macroscopic occupation of quasi-bound states is continuously replenished by the extraction of the black hole's rotational energy. This dynamic intrinsically favors emission over absorption once the resonance condition is met. Hence, robust stimulated emission proceeds efficiently without the need for a highly restrictive inverted state. The initial seeding required to trigger this macroscopic cascade can be naturally supplied either by an ambient stochastic gravitational-wave background or by spontaneous quantum fluctuations originating within the bosonic cloud itself. 

Translating these macroscopic beacons into observable signatures, the energy quantum $k_0$ released during a specific transition, such as the decay from the excited $|3, 2, 2\rangle$ state to the $|1, 0, 0\rangle$ ground state, is strictly governed by the hydrogenic fine-structure splitting: $k_{0}=\frac{4}{9}m_{b}c^{2}\alpha^{2}$. Spanning the relevant ultralight boson mass window of $m_{b}\in[10^{-58}, 10^{-46}]\, \text{kg}$, this single-particle energy release is infinitesimally small, bounded between $4\times10^{-44}\, \text{J}$ and $4\times10^{-32}\, \text{J}$. Crucially, however, the kinematics of this transition directly dictate the observable frequency of the coherent gravitational radiation. For a benchmark mass of $m_b = 10^{-46} \, \text{kg}$ at maximal superradiance ($\alpha = 0. 1$), the resonant emission peaks at $f \approx 241. 47\, \text{Hz}$ ($\lambda \approx 1. 2\times 10^6 \, \text{m}$). This aligns robustly with the peak sensitivity band of current ground-based interferometers like LIGO (10 Hz–10 kHz) \citep{Brito2015}. Conversely, extending the parameter space to lighter bosons naturally shifts the emission into the ultra-low-frequency domain, indicating that gravitational atoms can populate a wide array of observational bands spanning from audio to nanohertz frequencies. More fundamentally, because the emission is driven by discrete resonant transitions, the resulting radiation is exceptionally monochromatic. This persistent, narrowband spectral morphology differs fundamentally from the transient, broadband chirping typically associated with compact binary coalescences. Thus, the detection of a continuous, narrowly peaked gravitational-wave strain would constitute a qualitatively distinct signature of the quantized energy spectrum surrounding Kerr black holes, effectively functioning as an astrophysical spectrometer for ultralight bosonic fields. 

Dictated by the resonance condition $E_{nlm} - E_{n'l'm'} = E_0$, the characteristic frequency of the stimulated gravitational radiation depends critically on the underlying boson mass and the gravitational fine-structure constant. Consequently, the accessible frequency domain spans a vast astrophysical parameter space, potentially ranging from the $\sim 10\, \mathrm{pHz}$ regime up to the $\mathrm{kHz}$ band. This extensive coverage operates in a profoundly different regime from its electromagnetic counterpart, where terrestrial atomic transitions are heavily concentrated within the narrow infrared to ultraviolet window ($10^{13}$--$10^{16}\, \mathrm{Hz}$). The dichotomy between gravitational and electromagnetic stimulated emission extends fundamentally beyond their spectral footprints. In conventional quantum optical systems, bound electrons act as the active medium, transitioning under the influence of an external electromagnetic field. In the gravitational analogue, the gain medium is comprised of a macroscopic condensate of ultralight bosons, while ambient gravitational-waves provide the necessary resonant trigger. Furthermore, whereas atomic lasers depend on meticulously engineered cavities and artificial population inversions within controlled laboratories, gravitational atoms are spontaneously assembling astrophysical engines. Governed entirely by general relativistic dynamics, these natural amplifiers operate within the extreme, untamed spacetimes of rotating black holes, offering a radically different observational paradigm from human-calibrated instruments. 

In realistic gravitational atom systems, there typically exist not only the central black hole and ultralight bosons but also an accretion disk around the black hole. Consequently, the observed systems are generally complex. This raises the question: in such a complex realistic environment, is the observation of stimulated gravitational radiation feasible? Our answer is yes. On the one hand, the transitions of ultralight bosons emit gravitational-waves, and these bosons hardly interact with the Standard Model, which provides a decisive condition for their immunity to interference. Second, there is a realistic example: the maser phenomenon. 

\begin{table}[htpb]
\centering
\footnotesize
\setlength{\tabcolsep}{4pt}
\begin{tabular}{@{}lcc@{}}
\toprule
\textbf{Property} & \textbf{Astrophysical Maser} & \textbf{Gravitational Atom} \\
\midrule
\textbf{Working medium} 
& Atoms or molecules (e. g. , H$_2$O, OH) 
& Ultralight scalar bosons \\

\addlinespace[2pt]
\textbf{Radiated quanta} 
& Photons (microwaves) 
& gravitational-waves (Spin-2) \\

\addlinespace[2pt]
\textbf{Fundamental coupling} 
& Electromagnetic ($\sim e^2$) 
& Gravitational ($\sim \alpha^2$) \\

\addlinespace[2pt]
\textbf{Stimulating field} 
& Background photons / collisions 
& Stochastic GW background \\

\addlinespace[2pt]
\textbf{Population inversion} 
& Required (via thermal/radiative pumping) 
& Not strictly required (Bose enhanced) \\

\addlinespace[2pt]
\textbf{Typical frequency} 
& $1$--$100$\, GHz 
& $10^{-11}$--$10^{3}$\, Hz \\

\addlinespace[2pt]
\textbf{Signal character} 
& Narrow-band microwave lines 
& Persistent, narrow-band GW features \\
\bottomrule
\end{tabular}
\caption{Comparison of stimulated emission mechanisms in astrophysical masers and gravitational atoms. The table underscores the profoundly different parameter space of the gravitational analog, particularly noting its reliance on ultra-weak gravitational couplings, purely bosonic gain media, and the circumvention of strict population inversion. }
\label{tab:stimulated-emission-comparison}
\end{table}

Townes and collaborators in 1954, masers have illustrated how macroscopic amplification of radiation can arise through stimulated emission in a resonant gain medium sustained by a pumping mechanism and extended effective path lengths \citep{1954Gordon, Gordon1955}. Natural astrophysical masers (such as $H_{2}O$, $SiO$, and $OH$) further demonstrate that coherent amplification can emerge even in dilute, dynamic, and thermally active environments, provided resonance is maintained over sufficiently long distances\cite{Barrett1965}. While the maser analogy offers a useful conceptual guide, the underlying physical processes differ in several essential aspects. Conventional masers operate in the microwave band ($1-100\text{GHz}$), rely on strong electromagnetic coupling, and often require explicit population inversion together with feedback from cavities or extended coherent paths. 

An instructive phenomenological analogy is provided by astrophysical masers \citep{Elitzur1992, Gray2012}, which demonstrate that large-scale radiative amplification can arise naturally even in the absence of a well-defined optical resonant cavity. Nevertheless, as summarized in \ref{tab:stimulated-emission-comparison}, the physical mechanism underlying gravitational atoms differs fundamentally from its electromagnetic counterpart. One key distinction concerns the thermodynamic requirements necessary for amplification. Conventional masers require highly nontrivial environmental pumping mechanisms to sustain population inversion and overcome thermal absorption. Superradiant condensates, however, entirely bypass this bottleneck. Governed by Bose Einstein statistics, the macroscopic occupation of quasibound states is continuously replenished through the extraction of rotational energy from the black hole, thereby efficiently driving the emission process without the need for a finely tuned inverted population. 

\subsection{Microscopic Mechanism and Observability of Stimulated Emission from Gravitational Atoms}
To discuss the microscopic mechanism of gravitational atoms under the influence of a stochastic gravitational-wave background, it is useful to consider a simple example. We therefore focus on the transition of a particle initially occupying the bound state $|3, 2, 2\rangle$ into the final state $|1, 0, 0;\vec{k}, s=\pm2\rangle$. This process represents the simplest nontrivial example allowed by the transition selection rules. 

We begin with the stimulated emission process. The Universe is permeated by gravitational-waves emitted from a wide variety of astrophysical sources. The superposition of these waves forms an effectively stochastic gravitational-wave background containing modes across a continuous frequency spectrum. This provides one of the two necessary ingredients for stimulated transitions in gravitational atoms. The second ingredient is the existence of ultralight bosons generated through the superradiant extraction of rotational energy from Kerr black holes, which subsequently form gravitationally bound states around their parent black holes, namely gravitational atoms. Such systems may be regarded as gravitational analogues of the hydrogen atom, consisting of a central black hole surrounded by a cloud of ultralight bosons. 

Since the superradiance process extracts energy exclusively from the rotational energy reservoir of the Kerr black hole, the total number of produced bosons is necessarily finite and depends on the mass of the parent black hole. This follows from the superradiance condition, $\omega < m\Omega_H$, which imposes an inverse relationship between the black-hole mass and the boson mass. Consequently, more massive black holes preferentially generate lighter bosons, whereas lighter black holes favor heavier bosons. Current studies typically consider the boson mass range $m_b \in (10^{-58}, 10^{-46})\mathrm{kg}$, where the lower and upper limits correspond approximately to parent black-hole masses of $10^8 M_\odot$ and $1M_\odot$, respectively. Furthermore, it is generally believed that the total mass accumulated in the boson cloud through superradiance can reach roughly $10^{-1}$ of the black-hole mass. This implies a total boson occupation number in the range $N\in(10^{75}, 10^{99})$, with the lower bound corresponding to $m_b=10^{-46}\mathrm{kg}$ and the upper bound to $m_b=10^{-58}\mathrm{kg}$. Owing to the principle of minimum energy, the majority of bosons occupy low-lying bound states, thereby providing the large particle occupation numbers required for stimulated amplification induced by gravitational-waves. 

For the stimulated-emission scenario, let us assume that initially there are $N_{i}$ bosons with helicity $s=-2$ and momentum $k_{z}$ occupying the $|3, 2, 2\rangle$ state within the gravitational atom. The system is then perturbed by an incoming stochastic gravitational-wave. Among the continuous spectrum of incident gravitational-waves, there exist modes whose quantized frequencies precisely match the energy splitting associated with the transition from the $|3, 2, 2\rangle$ state to the ground state. These resonant modes induce transitions of bosons into the $|1, 0, 0\rangle$ state. During this process, each boson undergoing the transition emits an additional gravitational-wave identical to the incident mode. Macroscopically, the incoming gravitational-wave is therefore amplified, with the amplification factor proportional to the initial occupation number $N_{i}$, or more precisely $N_{i}+1$. 

In this sense, the stochastic gravitational-wave background acts as a highly selective quantum filter. Only those frequency components within the continuous spectrum that simultaneously satisfy both the energy resonance condition and the quadrupolar selection rules are preferentially amplified. This resonant filtering fundamentally shapes the emitted spectrum, ensuring that sharp and discrete emission lines emerge above the otherwise continuous stochastic background. 

Moreover, the phase coherence of these discrete emission lines is expected to be exceptionally well protected. In conventional astrophysical masers, phase coherence is ultimately degraded by collisional dephasing and thermal fluctuations within baryonic media. By contrast, ultralight bosons couple only negligibly to Standard Model particles, implying that the condensate remains almost completely isolated from environmental scattering even in dense and turbulent accretion disks. Combined with the extremely weak interaction cross section of gravitational-waves themselves, this isolation strongly suppresses collisional decoherence. Collectively, these dynamical advantages imply that the stimulated emission is not only highly spectrally peaked but also characterized by extraordinarily long coherence times, allowing such systems to behave as remarkably stable and long-lived gravitational-wave beacons. 

An important question is the magnitude of the amplification produced by stimulated emission. This is directly related to the decay properties of the gravitational atom and can therefore be understood by comparing the stimulated-emission process with the spontaneous decay of the $|3, 2, 2\rangle$ state. 

Our calculations show that the spontaneous emission rate is independent of the properties of any incident gravitational-wave. Instead, it is determined solely by the gravitational fine-structure constant $\alpha$, the ultralight boson mass $m_{b}$, and the occupation number of the bound state. In contrast, the stimulated emission rate depends not only on these same quantities, but also on the characteristic strain and frequency of the incoming gravitational-wave. 

This difference can be illustrated through a representative example. During the superradiance process around a Kerr black hole, the allowed parameter space constrains the gravitational fine-structure constant to values typically of order $10^{-1}$. It is therefore customary to adopt $\alpha \sim 0. 1$. Since spontaneous and stimulated emission coexist within the same gravitational-atom system, the numerical values of $\alpha$ and $m_{b}$ entering both processes are identical, allowing their common dependence to be factored out in a direct comparison. However, there is one crucial distinction: spontaneous emission does not exhibit coherent enhancement, whereas stimulated emission allows the emitted waves from particles occupying the same energy level to interfere coherently. Consequently, the stimulated signal acquires an additional enhancement proportional to the occupation number $N_{i}$. 

Furthermore, the stimulated process also depends on the gravitational-wave frequency and on the frequency-dependent characteristic strain of the stochastic gravitational-wave background. The transition energy condition implies that different boson masses correspond to different transition frequencies. We therefore consider a concrete benchmark example with $m_b = 4\times10^{-46}\mathrm{kg}$. For this choice, the transition from the $|3, 2, 2\rangle$ state to the ground state emits gravitational-waves with frequency $f = 241. 47\mathrm{Hz}$. Within the sensitivity range of current gravitational-wave detectors, characteristic strains around frequencies of several hundred hertz are typically of order $h_c \sim 10^{-24}$. We therefore adopt this representative value, from which we find that the stimulated emission is approximately $29$ to $31$ orders of magnitude stronger than the corresponding spontaneous emission. 

It should be emphasized that this does not imply that the effect is directly observable with current detector sensitivities, since the comparison above refers only to the difference in transition rates per unit time. To assess actual detectability, one must instead consider the resulting characteristic strain over astrophysical distances. Taking a representative propagation distance of one billion light-years, we find that the characteristic strain generated by stimulated emission exceeds that of spontaneous emission by roughly (16) orders of magnitude. Nevertheless, the resulting strain amplitude remains only of order $10^{-47}$, which is still far below the sensitivity of present-day gravitational-wave detectors. However, this does not imply that detecting gravitational-waves through stimulated emission in gravitational atoms is fundamentally impossible. The estimate presented here is based only on a stochastic gravitational-wave background model, for which the characteristic strain is relatively small. If instead one considers binary gravitational-atom systems, the characteristic strain could be enhanced substantially, potentially by many orders of magnitude. Consequently, the idea of gravitational-wave detection through stimulated emission remains a highly promising direction with considerable room for further theoretical development. 

In summary, the synthesis of immense macroscopic Bose enhancement, exceptional phase stability, and strict selection rules establishes gravitational atoms as potent, naturally occurring gravitational-wave amplifiers. Rather than relying on traditional matched-filtering templates tailored for compact binary coalescences, extracting these quantized emission features will require the development of dedicated continuous-wave data analysis pipelines. By mapping the discrete energy spectrum of the superradiant condensate, these stimulated emission signatures offer a pristine, dynamically rich observational framework for probing ultralight fields, constraining physics beyond the Standard Model, and exploring the intricate energy extraction mechanisms of rotating black holes \citep{Allen1999, Romano2017}. 

\section{Conclusion}\label{sec13}
In this work, we have established a theoretical framework demonstrating that Kerr black holes, when dressed by superradiant bosonic clouds, can act as natural amplifiers for gravitational-waves. While phenomenologically parallel to astrophysical masers, this mechanism operates fundamentally through macroscopic Bose enhancement and ultra-weak spacetime couplings, dispensing with the need for environmental pumping or explicit resonant cavities. 

By evaluating the transition dynamics within an S-matrix formalism, we found that stimulated emission driven by a stochastic gravitational-wave background heavily dominates over spontaneous vacuum decay, enhancing the transition rate by dozens of orders of magnitude. However, translating these linear transition rates into macroscopic observable strain reveals a significant dynamical threshold. The gravitational-wave strain generated strictly under this linear mechanism remains well below the sensitivity limits of current interferometers. Crucially, this limitation is a direct consequence of the linear perturbative regime. The physical realization of this mechanism implies that initially amplified waves could back-react on the condensate, seeding subsequent transitions. This introduces the potential for a non-linear, cascading avalanche of coherent radiation, which could boost the macroscopic strain significantly. Quantifying this runaway process requires a full dynamic treatment of the level populations, marking a primary and compelling objective for future numerical relativity investigations. 

If driven into the observable regime via such non-linearities, the resulting signals would offer a pristine observational target: persistent, highly monochromatic continuous gravitational-waves. Dictated by the underlying resonance conditions, these signatures naturally populate a broad frequency landscape depending on the boson mass. Benchmark masses around $m_b = 4. 0 \times 10^{-46}\, \text{kg}$ map to the audio band ($f \approx 241. 47\, \text{Hz}$) of ground-based detectors like LIGO and Virgo. Conversely, intermediate ($m_b = 4. 7 \times 10^{-51}\, \text{kg}$) and ultra-light ($m_b = 4. 7 \times 10^{-57}\, \text{kg}$) candidates align with the mHz band of LISA and the nHz window of pulsar timing arrays (e.g., NANOGrav, EPTA), respectively. This distinct narrowband morphology provides a clear discriminant against the transient, broadband chirps typically associated with compact binary coalescences. 

Ultimately, the primary contribution of this study is the formal identification of gravitational-wave stimulated amplification as a physically viable process in strong-gravity environments. Rather than presenting an immediately detectable source under linear assumptions, our findings establish a novel mechanism that could actively shape the continuous gravitational-wave sky. Future theoretical efforts focusing on non-linear saturation effects, combined with dedicated continuous-wave data analysis pipelines, will determine the ultimate detectability of these macroscopic quantum systems. This framework offers a rigorous new pathway to probe ultralight dark matter candidates and the near-horizon physics of rotating black holes. 

\section*{Acknowledgements}
 \quad We thank Rong-Gen Cai, Yu Tian, Qing-guo Huang and Shao-Feng Wu for helpful discussions. The work of XHG was partially supported by the National Natural Science Foundation of China under grant No. 12275166 and 12311540141. The work of YGG was partially supported by the National Natural Science Foundation of China under grant No. 12535002. The work of YLZ was partially supported by the National Natural Science Foundation of China under grant No. 12375059. 
\section{Supplementary Material}\label{secA0}

\subsection{Derivation of Bound State Wavefunctions and Energies}To rigorously quantify the stimulated transition rates, we must first establish the quantized field operator and the spatial morphology of the ultralight boson cloud. While the central Kerr black hole provides the necessary gravitational potential and angular momentum to dynamically grow the superradiant condensate, the kinematics of the stimulated emission are predominantly governed by the far-field monopole interactions. In the long-wavelength limit ($\alpha = M m_b \ll 1$), the bulk of the macroscopic condensate is localized at distances $r \sim \mathcal{O}(\alpha^{-2})M \gg M$, where the frame-dragging effects of the black hole spin are heavily suppressed. Consequently, the spatial wavefunctions and the real energy spectrum of the cloud are excellently approximated by a non-rotating, hydrogenic background. The black hole spin primarily dictates the superradiant instability rate (the imaginary part of the frequency) rather than the transition matrix elements. Thus, to isolate the fundamental physics of the stimulated emission, we treat the background geometry as effectively spherically symmetric in the spatial domain of the cloud. Assuming a real, massive scalar field, the dynamics of the ultralight bosons are governed by the Klein-Gordon equation in the Kerr spacetime. We promote the classical field to a Hermitian quantum operator ($\phi^\dagger = \phi$), expanding it in terms of the quasi-bound state wavefunctions:\begin{equation}\phi(t, \vec{r}) = \frac{1}{\sqrt{2 m_b}} \left[ \psi(t, \vec{r}) e^{-i m_b t} + \psi^*(t, \vec{r}) e^{i m_b t} \right]. \end{equation}Decomposing the positive-frequency mode as $\psi(t, \vec{r}) = e^{-i \omega t + i m \varphi} S(\theta) R(r)$ and substituting it into the Klein-Gordon equation yields the decoupled angular and radial equations \citep{Starobinskii1973, Brill1972, Rowan1976}:
\begin{align}
\frac{1}{\sin\theta} \frac{d}{d\theta} \left( \sin\theta \frac{d S(\theta)}{d\theta} \right) + \left[ a^2 (\omega^2 - m_b^2) \cos^2\theta - \frac{m^2}{\sin^2\theta} + \lambda \right] S(\theta) &= 0, \\
\Delta \frac{d}{dr} \left( \Delta \frac{d R}{dr} \right) + \left[ \omega^2 (r^2 + a^2)^2 - 4 a M r m \omega + a^2 m^2 - \Delta (m_b^2 r^2 + a^2 \omega^2 + \lambda) \right] R &= 0, \end{align}where $\Delta = r^2 - 2 M r + a^2$. Applying the long-wavelength ($\alpha \ll 1$) and far-field ($r \gg M$) approximations, the separation constant simplifies to $\lambda \approx l(l+1) + \mathcal{O}([a^{2}(\omega^{2}-m_{b}^{2})])$ \citep{Detweiler1980}. The equations subsequently reduce to:
\begin{align}\frac{d^2 (r R)}{dr^2} + \left[ \omega^2 - m_b^2 + \frac{2 M m_b^2}{r} - \frac{l(l+1)}{r^2} \right] r R &= 0, \\
\frac{1}{\sin\theta} \frac{d}{d\theta} \left( \sin\theta \frac{d S(\theta)}{d\theta} \right) + \left[ -\frac{m^2}{\sin^2\theta} + l(l+1) \right] S(\theta) &= 0. \end{align}This reduced system is isomorphic to the radial and angular components of the Schrödinger equation for a hydrogen-like atom:\begin{equation}i \frac{\partial}{\partial t} \psi(t, \vec{r}) = \left( -\frac{1}{2 m_b} \nabla^2 - \frac{\alpha}{r} \right) \psi(t, \vec{r}). \end{equation}The resulting quasi-bound state solutions are:\begin{equation}\psi_{nlm}(t, \vec{r}) = R_{nl}(r) Y_{lm}(\theta, \varphi) e^{-i (\omega_{nlm} - m_b) t}, \end{equation}where the angular distribution is governed by the standard spherical harmonics, \begin{equation}Y_{lm}(\theta, \varphi) = (-1)^m \sqrt{\frac{(2l+1)(l-m)!}{4 \pi (l+m)!}} P_l^m(\cos\theta) e^{i m \varphi}, \end{equation}and the radial profile is given by, \begin{equation}R_{nl}(r) = \sqrt{\left( \frac{2 m_b \alpha}{n} \right)^3 \frac{(n-l-1)!}{2 n (n+l)!}} \left( \frac{2 \alpha m_b r}{n} \right)^l e^{-\frac{m_b \alpha r}{n}} L_{n-l-1}^{2l+1} \left( \frac{2 \alpha m_b r}{n} \right), \end{equation}with $L_{n-l-1}^{2l+1}$ representing the generalized Laguerre polynomials, defined for $n \geq 1$, $l < n$, and $m = -l, \ldots, l$. The complex eigenfrequencies of these states are $\omega_{nlm} = E_{nlm} + i \Gamma_{nlm}$, where the real energy levels and the superradiant instability rates are, respectively:\begin{align}E_{nlm} &= m_b \left( 1 - \frac{\alpha^2}{2 n^2} + \cdots \right), \\\Gamma_{nlm} &= 2 r_+ C_{nl} g_{lm}(\tilde{a}, \alpha, \omega) (m \Omega_+ - \omega) \alpha^{4l+5} + \cdots. \end{align}Here, $\tilde{a} = a/M$ is the dimensionless black hole spin parameter, and the numerical coefficients are explicitly defined as:\begin{align}C_{nl} &= \frac{2^{4l+1} (n+l)!}{n^{2l+4} (n-l-1)!} \left[ \frac{l!}{(2l)! (2l+1)!} \right]^2, \\g_{lm} &= \prod_{k=1}^l \left[ k^2 (1 - \tilde{a}^2) + (\tilde{a} m - 2 r_+ \alpha)^2 \right]. \end{align}

Here, the real component $E_{nlm}$ represents the binding energy of the quasi-bound state, while the imaginary component $\Gamma_{nlm}$ quantifies the superradiant instability rate. The positive sign of $\Gamma_{nlm}$ dictates that the wave amplitude is governed by an exponentially growing temporal envelope, $e^{\Gamma_{nlm}t}$. Fundamentally, this macroscopic amplification is driven by the superradiant resonance condition, $\Gamma_{nlm} \propto (m\Omega_+ - \omega) > 0$. As the instability develops, the bosonic condensate dynamically extracts rotational energy and angular momentum from the black hole's ergoregion, leading to the rapid macroscopic growth of the cloud. Crucially, this exponential growth is not strictly unbounded; it is ultimately regulated by gravitational backreaction. As the cloud extracts angular momentum, the parent black hole progressively spins down, which in turn decreases the horizon angular velocity $\Omega_+$. This dynamic evolution continues until the system reaches superradiant saturation, defined by the threshold $\omega \approx m\Omega_+$. At this juncture, the energy extraction is quenched ($\Gamma_{nlm} \to 0$), and the system relaxes into a long-lived, quasi-stationary configuration. Rather than completely shedding its spin to become a non-rotating Schwarzschild black hole, the final state is a slower-spinning Kerr black hole enveloped by a maximally grown, stable superradiant condensate. It is precisely this incredibly stable, saturated macroscopic cloud that functions as the persistent gain medium for the subsequent stimulated emission of gravitational-waves. 

\subsection{Field Operator and Hamiltonian in the Interaction Picture}To rigorously compute the transition amplitudes induced by external gravitational-wave perturbations, we must transition to a second-quantized framework. Building upon the quasi-bound state basis derived above, we promote the scalar field to a quantum operator. Following standard canonical quantization \citep{Yoshino2012, Frolov1998}, the free field operator $\phi(t, \vec{r})$ and its conjugate momentum $\pi(t, \vec{r}) = \sqrt{-g}g^{0\mu}\partial_{\mu}\phi$ are expanded in terms of the discrete superradiant modes:\begin{equation}\phi(t, \vec{r}) = \sum_{nlm} \frac{1}{\sqrt{2E_{nlm}}}\left[ a_{nlm} \psi_{nlm}(\vec{r}) e^{-i E_{nlm} t} + a_{nlm}^{\dagger} \psi_{nlm}^*(\vec{r}) e^{i E_{nlm} t} \right], \end{equation}where $\psi_{nlm}(\vec{r}) = R_{nl}(r) Y_{lm}(\theta, \varphi)$ denotes the spatial component of the wavefunctions. The coefficients $a_{nlm}$ and $a_{nlm}^{\dagger}$ act as the canonical annihilation and creation operators for the state $|n, l, m\rangle$, satisfying the standard bosonic commutation relations:\begin{equation}[a_{nlm}, a_{n'l'm'}] = 0, \quad [a_{nlm}^{\dagger}, a_{n'l'm'}^{\dagger}] = 0, \quad [a_{nlm}, a_{n'l'm'}^{\dagger}] = \delta_{nn'} \delta_{ll'} \delta_{mm'}. \end{equation}Operating in the weakly interacting regime, the unperturbed dynamics of the macroscopic boson cloud are governed by the free Hamiltonian, $H_B$, which naturally assumes a diagonalized form in this eigenbasis:\begin{equation}H_B = \sum_{nlm} E_{nlm} a_{nlm}^{\dagger} a_{nlm}, \end{equation}where the zero-point vacuum energy $E^{(0)}$ has been conventionally renormalized to zero. To systematically treat the scattering and stimulated emission processes via the S-matrix formalism, we shift to the interaction picture. Driven exclusively by the unperturbed Hamiltonian $H_B$, the time evolution of the field operator in the interaction picture, $\phi_I(t, \vec{r}) = e^{i H_B t} \phi_S(\vec{r}) e^{-i H_B t}$, explicitly retains the mode expansion structure:\begin{equation}\phi_I(t, \vec{r}) = \sum_{nlm} \frac{1}{\sqrt{2E_{nlm}}} \left[ a_{nlm} \psi_{nlm}(\vec{r}) e^{-i E_{nlm} t} + a_{nlm}^{\dagger} \psi_{nlm}^*(\vec{r}) e^{i E_{nlm} t} \right]. \end{equation}This explicit formulation of $\phi_I(t, \vec{r})$ provides the fundamental quantum infrastructure required to evaluate the interaction Hamiltonian and derive the transition matrix elements rigorously \citep{Peskin1995}. 

\subsection{Quantized Gravitational-Wave Field}To evaluate the quantum transition amplitudes driven by the stochastic gravitational-wave background, we treat the ambient gravitational radiation within the framework of linearized effective field theory. In the local weak-field regime, the dynamics of the metric perturbation $h_{\mu\nu}$ on a flat background are governed by the Fierz-Pauli Lagrangian density \citep{Fierz1939}. By adopting the transverse-traceless gauge, wherein $\nabla_\mu h^{\mu\nu} = 0$ and $h = 0$, the Lagrangian elegantly reduces to the canonical kinetic term for a massless spin-2 field:\begin{equation}\mathcal{L} = -\frac{1}{2} \nabla_\alpha h_{\mu\nu} \nabla^\alpha h^{\mu\nu}. \end{equation}This simplified free-field action allows us to promote the classical metric perturbation to a quantum field operator, expanding it in terms of plane-wave momentum modes. Because the stimulated transition rules ($\Delta m = \pm 2$) are fundamentally dictated by angular momentum transfer, it is most physically transparent to express the field in the circular helicity basis ($\lambda = \pm 2$). These helicity tensors are related to the standard linear polarization modes by $e_{\mu\nu}(\vec{k}, \pm 2) = \frac{1}{\sqrt{2}} \left[e_{\mu\nu}(\vec{k}, +) \pm i e_{\mu\nu}(\vec{k}, \times)\right]$. 

Normalizing the field to ensure the standard canonical commutation relations, the graviton field operator in natural units expands as \citep{Weinberg1972}:\begin{equation}h_{\mu\nu}(x) =  \sum_{\lambda=\pm 2} \int \frac{d^3k\sqrt{16\pi G_N}}{(2\pi)^{3/2}\sqrt{2\omega_k}} \left[ a(\vec{k}, \lambda) e_{\mu\nu}(\vec{k}, \lambda) e^{ik\cdot x} + a^\dagger(\vec{k}, \lambda) e_{\mu\nu}^*(\vec{k}, \lambda) e^{-ik\cdot x} \right], \end{equation}where $k = (\omega_k, \vec{k})$ represents the null four-momentum with angular frequency $\omega_k = |\vec{k}|$. The coefficients $a(\vec{k}, \lambda)$ and $a^\dagger(\vec{k}, \lambda)$ denote the annihilation and creation operators for a graviton with wave vector $\vec{k}$ and helicity $\lambda$. They satisfy the standard bosonic commutation relations:\begin{equation}[a(\vec{k}, \lambda), a^\dagger(\vec{k}', \lambda')] = \delta^3(\vec{k} - \vec{k}') \delta_{\lambda\lambda'}, \quad [a, a] = 0, \quad [a^\dagger, a^\dagger] = 0. \end{equation}
Analogous to the scalar field treatment, this mode expansion represents the freely evolving field operator in the interaction picture. It provides the essential tensorial building blocks required to couple the external SGWB to the superradiant boson cloud in the subsequent S-matrix calculations. 
\subsection{Calculation of the stimulated emission transition rate from $|3, 2, 2;N_{i}(k_{z}, -2)\rangle$ to $|1, 0, 0;N_{i}(k_{z}, -2)+1\rangle$. }The fundamental coupling between the ultralight boson cloud and the stochastic gravitational-wave background (SGWB) is governed by the minimal coupling of the scalar field to the spacetime metric. Expanding the standard action for a massive scalar field, \begin{equation}S_{\phi} = -\frac{1}{2}\int d^4x \sqrt{-g} \left( g^{\mu\nu}\partial_\mu\phi \partial_\nu\phi + m_b^2 \phi^2 \right), \end{equation} to linear order in the metric perturbation $h_{\mu\nu}$, the interaction isolates the coupling to the scalar energy-momentum tensor. By imposing the transverse-traceless (TT) gauge ($h^\mu_\mu = 0$, $\partial^\mu h_{\mu\nu} = 0$), the trace components vanish rigorously. Operating in natural units, the leading-order interaction Lagrangian density thus simplifies to:
\begin{equation}\mathcal{L}_I = \frac{1}{2} h_{\mu\nu} \partial^\mu \phi \partial^\nu \phi. \end{equation}
Transitioning to the Hamiltonian formalism in the long-wavelength limit where the background Kerr geometry across the cloud is effectively treated as locally flat the corresponding interaction Hamiltonian density is simply given by $\mathcal{H}_I = -\frac{1}{2} h_{\mu\nu} \partial^\mu \phi \partial^\nu \phi$. To evaluate the stimulated emission rate quantitatively, we compute the transition matrix elements via the S-matrix Dyson series expansion. We define the initial composite state as $|i\rangle = |n, l, m\rangle \otimes |N_{\vec{k}, \lambda}\rangle$, representing a boson in an excited quasi-bound state immersed in an SGWB containing $N$ gravitons with wave vector $\vec{k}$ and helicity $\lambda$. The corresponding final state, following the stimulated emission of a coherent graviton, is $|f\rangle = |n', l', m'\rangle \otimes |N_{\vec{k}, \lambda} + 1\rangle$. Applying Wick’s theorem to the first-order transition amplitude, non-vanishing matrix elements are strictly governed by quadrupolar selection rules. For instance, considering an incident gravitational-wave propagating along the quantization axis ($\vec{k} = k\hat{z}$), the decay from the dominant superradiant excited state $|3, 2, 2\rangle$ to the ground state $|1, 0, 0\rangle$ selectively couples exclusively to the emission of a left-circularly polarized graviton ($\lambda = -2$). The resulting macroscopic transition rate, formally derived via Fermi’s golden rule, is ultimately weighted by the phase-space spectral density of the SGWB, integrating over the available density of states for the background gravitons. 

To quantitatively evaluate the stimulated emission rate, we compute the transition amplitude in the interaction picture. Under the adiabatic approximation, the leading-order transition between distinct bound-state energy eigenstates is governed by the first-order S-matrix term, $iT^{(1)} = -i \int d^{4}x \, \mathcal{T}[\mathcal{H}^{I}(x)]$, where $\mathcal{H}^{I}(x) = -\frac{1}{2} h_{\mu\nu}\partial^{\mu}\phi\partial^{\nu}\phi$. 

Applying Wick's theorem, we extract the connected normal-ordered operator containing one explicit $h_{\mu\nu}$ insertion, dropping disconnected vacuum bubbles in accordance with standard LSZ reduction. We define the initial state as $|i\rangle \equiv |3, 2, 2; N_{i}(k_{z}, -2)\rangle$ and the final state as $|f\rangle \equiv |1, 0, 0; N_{i}(k_{z}, -2)+1\rangle$, representing the stimulated emission of a left-circularly polarized graviton propagating along the $+z$ axis. For notational simplicity, we omit the first-order superscript in what follows. The transition matrix element is given by:
\begin{align}\nonumber
    T_{i \to f} &= \langle 1, 0, 0; N_{i}(k_{z}, -2)+1 | \frac{i}{2} \int d^{4}x \mathcal{N}[h_{\mu\nu}\partial^{\mu}\phi\partial^{\nu}\phi] | 3, 2, 2; N_{i}(k_{z}, -2) \rangle \\\nonumber
    &= \sqrt{16\pi G_{N}} \frac{i \sqrt{N_{i}+1}}{16\pi^{3/2} \sqrt{2|k_{z}|}} \frac{1}{\sqrt{4E_{322}E_{100}}} \int d^{4}x \, e^{-ik\cdot x} \\\nonumber
    &\quad \times \bigg[ \partial^{x}\psi^{*}_{100}\partial^{x}\psi_{322} - \partial^{y}\psi^{*}_{100}\partial^{y}\psi_{322} + i\partial^{x}\psi^{*}_{100}\partial^{y}\psi_{322} + i\partial^{y}\psi^{*}_{100}\partial^{x}\psi_{322} \bigg], 
\end{align}
where \(\mathcal{N}\) represents the normal ordering operation. The temporal integration naturally enforces energy conservation:
\begin{equation}
    \int_{-\infty}^{\infty} dt \, e^{i(k_{0}+E_{100}-E_{322})t} = 2\pi\delta(E_{322}-k_{0}-E_{100}). 
\end{equation}

To facilitate the subsequent calculations, we write the spatial integral as \(I = I_{+} + i I_{\times}\), where:
\begin{align}
    I_{+} &= \int d^{3}x \, e^{-ik_{z}z} \big( \partial^{x}\psi^{*}_{100}\partial^{x}\psi_{322} - \partial^{y}\psi^{*}_{100}\partial^{y}\psi_{322} \big), \\
    I_{\times} &= \int d^{3}x \, e^{-ik_{z}z} \big( \partial^{x}\psi^{*}_{100}\partial^{y}\psi_{322} + \partial^{y}\psi^{*}_{100}\partial^{x}\psi_{322} \big). 
\end{align}
Thus, the matrix element simplifies to:
\begin{equation}
    T_{i \to f} \equiv \sqrt{16\pi G_{N}} \frac{i \sqrt{N_{i}+1}}{8\pi^{1/2}} \frac{1}{\sqrt{2|k_{z}|}} \frac{\delta(E_{322}-k_{0}-E_{100})}{\sqrt{4E_{322}E_{100}}} (I_{+} + iI_{\times}). 
\end{equation}
Using the explicit hydrogenic wavefunctions, 
\begin{align}
    \psi_{100} &= \sqrt{\frac{(m_{b}\alpha)^{3}}{\pi}} e^{-m_{b}\alpha r}, \\
    \psi_{322} &= \sqrt{\frac{(m_{b}\alpha)^{3}}{\pi}} \frac{(m_{b}\alpha)^{2}}{162} r^{2} e^{-\frac{m_{b}\alpha r}{3}} \sin^{2}\theta e^{i2\varphi} \equiv \sqrt{\frac{(m_{b}\alpha)^{3}}{\pi}} \frac{(m_{b}\alpha)^{2}}{162} u_{32}(r) b_{22}(\theta, \varphi), 
\end{align}
we transform the Cartesian derivatives into spherical coordinates. Due to the azimuthal symmetry, the $\varphi$-dependent integrals evaluate straightforwardly to $\pi$ and $i\pi/2$. Evaluating the remaining radial and polar integrals for $I_{+}$, we find:
\begin{align}\nonumber
    I_{+} &= -\frac{2(m_{b}\alpha)^{6}}{81} \int_{0}^{\infty} dr \int_{0}^{\pi} d\theta \, r^{3}\sin^{3}\theta e^{-ik_{z}r\cos\theta} e^{-\frac{4m_{b}\alpha r}{3}} \\
    &\quad + \frac{(m_{b}\alpha)^{7}}{486} \int_{0}^{\infty} dr \int_{0}^{\pi} d\theta \, r^{4}\sin^{5}\theta e^{-ik_{z}r\cos\theta} e^{-\frac{4m_{b}\alpha r}{3}}. 
\end{align}
Integrating over $\theta$ yields combinations of spherical Bessel-like functions. Subsequent integration over $r$ provides exact rational and arctangent terms:
\begin{align}\nonumber
    I_{+} &= \frac{8(m_{b}\alpha)^{6}}{81k_{z}^{5}} \bigg[ -\frac{12m_{b}^{2}\alpha^{2}k_{z}(15k_{z}^{2}+16m_{b}^{2}\alpha^{2})+162k_{z}^{5}}{(9k_{z}^{2} + 16\alpha^{2}m_{b}^{2})^{2}} + m_{b}\alpha \arctan\left(\frac{3 k_z}{4 \alpha m_b}\right) \bigg]. 
\end{align}
A parallel computation for the cross-polarization integral confirms that $I_{+} = iI_{\times}$. 

To extract the invariant transition amplitude $\mathcal{M}_{i \to f}$ (factoring out the $2\pi\delta(E_{i}-E_{f})$ momentum-conserving delta function), we impose the superradiant resonance condition. For emission along the $+z$ axis, the graviton momentum satisfies $k_z = \omega_k = E_{322} - E_{100} \approx \frac{4}{9} m_b \alpha^2$. Substituting this resonant wavevector into our expression for $I_+$, we explicitly factorize the invariant amplitude as:
\begin{align}
       \mathcal{M}_{i \to f} &= - \sqrt{16\pi G_{N}} \frac{i \sqrt{N_{i}+1}}{16\pi^{3/2}} \frac{m_b^2}{\sqrt{4E_{322}E_{100}}}\\
       &\times\frac{9^{5}}{4^{5}} \frac{8}{81}\frac{1}{\alpha^{3}} \bigg[ \frac{\frac{1}{3}(\frac{5}{27}\alpha^{4}+\alpha^{2})+\frac{8}{9^{3}}\alpha^{6}}{(1+\frac{\alpha^{2}}{9})^{2}} - \alpha \arctan\left(\frac{\alpha}{3}\right) \bigg]. 
\end{align}
The macroscopic transition rate is formally dictated by Fermi's golden rule:
\begin{equation}
    \Gamma_{i \to f} = 2\pi \sum_{\lambda=\pm2} \int\frac{d^{3}k}{(2\pi)^{3}} \left| \mathcal{M}_{i \to f} \right|^2 \, \delta(E_{322} - E_{100} - \omega_k). 
\end{equation}
Substituting our derived matrix element and restoring SI units ($c$ and $\hbar$), the transition rate becomes:
\begin{equation}
    \Gamma_{i \to f} = \frac{G_{N}(N_{i}+1)}{256\pi^{4} E_{322} E_{100}} F^{2}(\alpha) m_{b}^{5} \frac{c}{\hbar^2},
\end{equation}
where the dimensionless shape function $F(\alpha)$ is analytically defined as:
\begin{equation}
    F(\alpha) \equiv \frac{9^{5}}{4^{5}} \frac{8}{81} \frac{1}{\alpha^{2}} \bigg[ \frac{\frac{1}{3}(\frac{5}{27}\alpha^{4}+\alpha^{2})+\frac{8}{9^{3}}\alpha^{6}}{(1+\frac{\alpha^{2}}{9})^{2}} - \alpha \arctan\left(\frac{\alpha}{3}\right) \bigg]. 
\end{equation}
In the maximal superradiance regime ($\alpha = 0. 1$), the shape function evaluates to $F(0. 1) \approx 1. 89774$. Using $E_{nlm} \approx m_b c^2$, the final transition rate takes the numerically evaluated form:
\begin{equation}
    \Gamma_{|3, 2, 2\rangle \to |1, 0, 0\rangle} \approx 2. 6 \times 10^{62} (N_{i}+1) m_{b}^{3} \, \mathrm{s}^{-1}. 
\end{equation}

Meanwhile, using the transition energy condition \(E_{322} - E_{100} = k_0\) and the ultralight boson mass range \(m_b \in (10^{-58}, 10^{-46})\ \text{kg}\), the energy range for a single particle transition can be obtained as:
\begin{align}
   k_{0}=\frac{4}{9}m_{b}c^{2}\alpha^{2}\in(4\times10^{-44}, 4\times10^{-32})\text{J}. 
\end{align}
On the other hand, considering the gravitational-wave quantization condition, the frequency range of gravitational-waves released by the stimulated emission process can be obtained as:
\begin{align}
   k_{0}=\hbar\omega\Rightarrow f\in(6\times10^{-11}, 60)\text{Hz}. 
\end{align}
\subsection{Calculation of $|322\rangle$ spontaneous emission}
To evaluate the total spontaneous decay rate of the dominant superradiant state, we must sum the partial transition rates to all kinematically allowed lower-lying energy levels. For the initial $|3, 2, 2\rangle$ state, the available phase space includes transitions to the following bound states:$$|2, 1, 1\rangle, \quad |2, 1, -1\rangle, \quad |2, 1, 0\rangle, \quad |2, 0, 0\rangle, \quad |1, 0, 0\rangle. $$The total spontaneous emission rate is thus the superposition of these partial channels:$$\Gamma_{|3, 2, 2\rangle} = \sum_{E_{f} < E_{322}} \Gamma_{|3, 2, 2\rangle \to |f\rangle}. $$While multiple decay channels exist, the tensor nature of the gravitational-wave perturbation imposes stringent angular momentum and parity selection rules. Because gravitons are massless spin-2 particles, transitions altering the orbital angular momentum by $\Delta l = 1$ (such as decays to the $|2, 1, m\rangle$ states) require parity-changing magnetic quadrupolar interactions, which are heavily suppressed in the non-relativistic limit. Consequently, the total decay rate is overwhelmingly dominated by the electric-quadrupolar transitions to the $l=0$ states ($|2, 0, 0\rangle$ and $|1, 0, 0\rangle$). To avoid exhaustive and redundant numerical integration for each sub-dominant channel, we can robustly extract the universal order of magnitude for the transition rates via dimensional analysis. For any final state $|f\rangle$, the leading-order transition matrix element takes the general S-matrix form:$$T^{(1)}_{i \to f; \vec{k}, \lambda} = i \sqrt{8\pi G_N} \frac{1}{\sqrt{\omega_k}} \frac{1}{\sqrt{4E_i E_f}} \delta(E_i - E_f - \omega_k) \int d^3x \, e_{ab}^*(\vec{k}, \lambda) e^{-i\vec{k}\cdot\vec{x}} \partial^a \psi_f^* \partial^b \psi_i. $$Let us define the core spatial overlap integral as $\mathcal{I}$. For these quasi-bound states, the characteristic spatial scale is dictated by the Bohr-like radius $r \sim (m_b\alpha)^{-1}$, which establishes the integration volume $\int d^3x \sim (m_b\alpha)^{-3}$. The normalized wavefunctions and their gradients dimensionally scale as $\psi \sim (m_b\alpha)^{3/2}$ and $\partial\psi \sim (m_b\alpha)^{5/2}$, respectively. Incorporating the order-unity polarization tensors and the long-wavelength approximation ($e^{-i\vec{k}\cdot\vec{x}} \sim \mathcal{O}(1)$), the spatial overlap integral scales universally as:$$\mathcal{I} \sim (m_b\alpha)^{-3} \big[ (m_b\alpha)^{5/2} \big]^2 = (m_b\alpha)^2. $$By factoring out the energy-conserving delta function to isolate the invariant amplitude $\mathcal{M}_{i \to f}$, the transition rate per unit time is derived via Fermi's golden rule:$$\Gamma_{i \to f} = 2\pi \sum_{\lambda=\pm2} \int \frac{d^3k}{(2\pi)^3} \delta(\Delta E - \omega_k) |\mathcal{M}_{i \to f}|^2 \sim \omega_k^2 |\mathcal{M}_{i \to f}|^2. $$Substituting the dimensional scaling of the amplitude ($|\mathcal{M}|^2 \sim \frac{G_N}{\omega_k E_i E_f} \mathcal{I}^2$), and recognizing that $E_i \approx E_f \approx m_b$, the transition rate simplifies to:$$\Gamma_{i \to f} \sim \omega_k^2 \left( \frac{G_N}{\omega_k m_b^2} \right) (m_b\alpha)^4 = \omega_k \frac{G_N}{m_b^2} (m_b\alpha)^4. $$Imposing the macroscopic resonance condition $\omega_k \sim m_b\alpha^2$, we find that all kinematically allowed primary transitions share the same universal parametric scaling. Thus, the total spontaneous emission rate is cleanly bounded by:$$\Gamma_{\text{sp}} \sim G_N m_b^3 \alpha^6. $$To evaluate this macroscopically, we restore the SI constants ($c$ and $\hbar$). Evaluating the rate at the maximal superradiance efficiency ($\alpha = 0. 1$), the expression yields:$$\Gamma_{\text{sp}} \sim m_b^3 c \alpha^6 \frac{G_N}{\hbar^2} \approx 1. 8 \times 10^{60} \, m_b^3 \, \text{s}^{-1}. $$Given the target mass window for ultralight scalar bosons $m_b \in (10^{-58}, 10^{-46}) \, \text{kg}$, the spontaneous emission rate falls within the exceedingly faint regime of:$$\Gamma_{\text{sp}} \sim (1. 8 \times 10^{-114}, \ 1. 8 \times 10^{-78}) \, \text{s}^{-1}. $$

It should be particularly noted that the calculations concerning spontaneous emission are only rough estimates, so slight deviations in the order of magnitude may occur. This is mainly because we need to consider and sum over all transitions from state $|322\rangle$ to lower energy levels. However, for the purpose of order-of-magnitude estimation, we combined all possible transitions and approximated the result using the order of magnitude of a single term instead of performing the full complex calculation, which may lead to variations in the resulting order of magnitude. 

\subsection{Selection Rules and the Stochastic Background Spectrum}

The non-vanishing character of the spatial overlap integral $\mathcal{M}_{i \to f}$ is fundamentally dictated by rigorous angular momentum selection rules. Because the incident gravitational perturbation is a transverse-traceless rank-2 tensor, its angular dependence expands naturally into spin-weighted spherical harmonics. Consequently, the transition matrix elements are governed by the Wigner-Eckart theorem, requiring the magnetic quantum numbers to satisfy $\Delta m = m' - m = q$, where $q \in \{-2, -1, 0, 1, 2\}$. For a pure helicity mode ($\lambda = \pm 2$) propagating along the quantization $z$-axis, the interaction exclusively selects the $q = \lambda$ component, restricting the allowed transitions to the strict quadrupolar rule:
\begin{align}
    \Delta m = \pm 2. 
\end{align}
This rigorous selection directly justifies our specific focus on the $|3, 2, 2\rangle \to |1, 0, 0\rangle$ channel triggered by a left-circularly polarized ($\lambda = -2$) graviton. 

While the S-matrix formalism yields the microscopic transition amplitude for a deterministic plane wave, coupling this system to an astrophysical stochastic gravitational-wave background requires integrating over the available phase space. In a realistic astrophysical environment, the idealized energy-conserving delta function is broadened into a physical line-shape profile, $g(\omega - \omega_0)$, characterized by an effective resonance window $\Delta\omega \simeq \max\{\Gamma_{\rm line}, T_{\rm coh}^{-1}\}$, where $\Gamma_{\rm line}$ is the intrinsic linewidth and $T_{\rm coh}$ is the coherence time of the background. 

To bridge the microscopic field-theoretic calculation with macroscopic astrophysical observables, it is highly instructive to employ the Einstein-coefficient framework. For a fixed bound-bound transition with resonant angular frequency $\omega_0 = (E_{322}-E_{100})/\hbar$, the spontaneous emission rate corresponds to the Einstein $A$ coefficient ($\Gamma_{\rm spont} \equiv A_{ji}$). Immersed in an incoherent SGWB with a spectral energy density $\rho_{\rm GW}(\omega)$, the stimulated rate is driven by the Einstein $B$ coefficient:
\begin{equation}
    \Gamma_{\rm stim} = B_{ji}\int d\omega\;\rho_{\rm GW}(\omega)\, g(\omega-\omega_0) \simeq B_{ji}\rho_{\rm GW}(\omega_0). 
\end{equation}
For a massless spin-2 bosonic field, the universal detailed-balance relation yields $A_{ji}/B_{ji} = \hbar\omega_0^{3}/(\pi^{2}c^{3})$. Consequently, the stimulated-to-spontaneous ratio is independent of the explicit wavefunctions and is fixed solely by the ambient radiation density at resonance:
\begin{equation}
    \frac{\Gamma_{\rm stim}}{\Gamma_{\rm spont}} \simeq \frac{\pi^{2}c^{3}}{\hbar\omega_0^{3}} \rho_{\rm GW}(\omega_0)\equiv \bar{n}_{eff}. 
\end{equation}
Here, $\bar{n}_{\rm eff}$ is the average free graviton occupation number of the radiation mode and precisely equals $N_i$ in the S-matrix field theory derivation.

In gravitational-wave astronomy, the SGWB is conventionally parameterized by the one-sided strain power spectral density $S_h(f)$, or equivalently the characteristic strain $h_c(f)$, defined via $h_c^2(f) \equiv f S_h(f)$. Consider the existence of relation $\Omega_{GW}(f)=\frac{1}{\rho_{c}}\frac{d\rho_{GW}}{dln^{f}}$, so using the standard cosmological relation $\Omega_{\rm GW}(f) = \frac{2\pi^{2}}{3H_0^{2}}f^{3}S_h(f)$ and the critical density $\rho_c = \frac{3H_0^{2}c^{2}}{8\pi G_N}$, where $H_{0}$ is Hubble constant and \(\Omega_{\text{GW}}(f)\) is the gravitational-wave energy density spectrum, the spectral energy density transforms as $\rho_{\rm GW}(\omega_0) = \frac{c^{2}}{8G_N} f_0 S_h(f_0)$. Substituting this into the Einstein ratio, we arrive at the pivotal relationship governing the macroscopic amplification:
\begin{equation}
    \frac{\Gamma_{\rm stim}}{\Gamma_{\rm spont}} \simeq \frac{c^{5}}{64\pi G_N \hbar} \frac{S_h(f_0)}{f_0} = \frac{c^{5}}{64\pi G_N \hbar} \frac{h_c^{2}(f_0)}{f_0^{2}}. 
\end{equation}

For the benchmark parameter space of $m_b = 4\times10^{-46}\, {\rm kg}$ and $\alpha = 0. 1$, the resonant frequency evaluates to $f_0 \simeq 241. 47\, {\rm Hz}$. Assuming a representative, conservative SGWB characteristic strain of $h_c(f_0) \sim 10^{-24}$ in the audio band, the enhancement factor reaches:
\begin{equation}
    \frac{\Gamma_{\rm stim}}{\Gamma_{\rm spont}} \simeq \frac{c^{5}}{64\pi G_N \hbar f_0^{2}} h_c^{2}(f_0) \approx 2. 9\times 10^{31}. 
\end{equation}
This staggering 31-order-of-magnitude enhancement elucidates a profound physical reality: because low-frequency gravitational radiation corresponds to infinitesimally small energy quanta, even an observationally faint strain background ($h_c \sim 10^{-24}$) harbors an astronomically immense the average free graviton occupation number ($\bar{n}_{\rm eff} \sim 10^{31}$). This collective macroscopic coherence robustly guarantees that, once the superradiant condensate is established, the transition dynamics are overwhelmingly dominated by stimulated emission. 

\subsection{Estimation of the Gravitational-Wave Strain Amplitude}

While the Einstein coefficient framework elegantly establishes the overwhelming transition rate dominance of stimulated emission ($\Gamma_{\rm stim}/\Gamma_{\rm sp} \sim 10^{31}$), evaluating the astrophysical detectability requires mapping these local decay rates to far-field gravitational-wave strain amplitudes. 

Please pay special attention here: in the following discussion, we use \(\Gamma_{\rm stim}/\Gamma_{\rm sp} \sim 10^{31}\) rather than the result obtained from the quantum field theory calculation presented earlier. The reason is that the spontaneous emission rate in the quantum field theory calculation is an estimated value, which may differ from the true result by a very small order of magnitude. Therefore, in the calculation here, we adopt this conclusion. It is worth mentioning that although there is a discrepancy of two orders of magnitude between the two conclusions, this difference can be fully attributed to the uncertainty inherent in the estimation. 

We first establish the conservative baseline by computing the characteristic strain $h_{\rm sp}(r)$ generated exclusively by spontaneous emission. Assuming the spontaneous decays are fundamentally incoherent across the superradiant cloud, the total spontaneous luminosity is given by $L_{\rm sp} \simeq N_{\rm exc} \hbar \omega_0 \Gamma_{\rm sp}$, where $N_{\rm exc}$ denotes the macroscopic occupation number of the parent quasi-bound state. In standard thermal systems, population is governed by energy minimization; however, superradiant condensates are dynamically driven. Because the $|3, 2, 2\rangle$ mode typically exhibits the maximum superradiant instability growth rate, it effectively sequesters the bulk of the rotational energy extracted from the Kerr black hole. Consequently, for the purpose of establishing a robust theoretical upper bound, we approximate the excited state population as the total cloud particle number: $N_{\rm exc} \approx N_{\rm total}$. 

For a source located at a luminosity distance of $r = 10^9\, \text{ly}$ ($\sim 300\, \text{Mpc}$), the spontaneous characteristic strain is approximated by the standard far-field quadrupolar flux limit:
\begin{equation}
h_{\rm sp}(r) \simeq \frac{1}{r}\sqrt{\frac{G_N\, L_{\rm sp}}{c^{3}\omega_0^{2}}}, 
\label{eq:hFromLuminosity_sp}
\end{equation}
up to order-one angular factors dictated by the system's beaming and polarization geometry. Substituting our derived benchmark values for the $m_b = 10^{-46}\, \text{kg}$ candidate ($\Gamma_{\rm sp} \sim 1. 8 \times 10^{-78}\, \text{s}^{-1}$, $N_{\rm total} = 10^{75}$), the spontaneous luminosity evaluates to an astronomically feeble level. Consequently, the baseline spontaneous strain is deeply sub-threshold, yielding:
\begin{equation}
h_{\rm sp}(r) \sim 3. 6 \times 10^{-63}. 
\label{eq:h_sp_value}
\end{equation}
A parallel evaluation for the lighter candidate ($m_b = 10^{-58}\, \text{kg}$, $N_{\rm total} = 10^{99}$) results in an identical order of magnitude, confirming that spontaneous emission from these systems is universally unobservable. 

The stimulated emission channel, however, radically alters this output. Treating the stimulated process conventionally as an enhanced effective luminosity, $L_{\rm st} \simeq (\Gamma_{\rm stim}/\Gamma_{\rm sp}) L_{\rm sp}$, the observable strain scales precisely as the square root of the transition rate ratio. Leveraging our previously derived macroscopic enhancement factor, the stimulated characteristic strain escalates by 16 orders of magnitude:
\begin{equation}
h_{\rm st}(r) \simeq h_{\rm sp}(r) \sqrt{\frac{\Gamma_{\rm stim}}{\Gamma_{\rm sp}}} \sim 3. 6 \times 10^{-47}. 
\label{eq:h_st_value}
\end{equation}

While this mathematically demonstrates a colossal $10^{16}$ relative amplification in amplitude, treating stimulated emission merely as an isotropic power source represents a highly conservative geometric approximation. In physical reality, stimulated gravitons are strictly phase aligned and highly directional, inheriting the precise wavevector of the incident seed wave. Therefore, the true phenomenological observable is more robustly characterized as a directional gain coefficient amplifying the specific resonant modes of the incident stochastic background along the effective coherence path of the cloud rather than an isotropic spherical emission. Nevertheless, this 16-order-of-magnitude enhancement establishes the fundamental efficacy of the mechanism, explicitly underscoring the critical necessity of incorporating stimulated interactions when modeling the gravitational-wave phenomenology of superradiant black hole systems. 
\bibliography{sn-bibliographyGRlaser}
\end{document}